\documentstyle[aps,prl,epsf,multicol]{revtex}

\begin{document}
\title{Stress Propagation in Two Dimensional Frictional Granular Matter}
\author{Rava da Silveira}
\address{
Lyman Laboratory of Physics, Harvard University, Cambridge, Massachusetts
02138, U.S.A. \\
rava@physics.harvard.edu}
\author{Guillaume Vidalenc and Cyprien Gay}
\address{
Centre de Recherche Paul Pascal --- CNRS, Avenue A. Schweitzer, 33600
Pessac, France \\
vidalenc@crpp.u-bordeaux.fr, cgay@crpp.u-bordeaux.fr }

\maketitle

\begin{abstract}
The stress profile and reorientation of grains, in response to a point force
applied to a preloaded two dimensional granular system, are calculated in
the context of a continuum theory that incorporates the texture of the
packing. When high friction prevents slip at the inter-grain contacts, an
anisotropic packing propagates stress along two peaks which amalgamate into
a single peak as the packing is disordered into a less anisotropic
structure; this single peak may be wider or narrower than in the isotropic
case, depending on the preparation of the packing. At lower frictions, an
effective treatment of slipping contacts yields sharpened peaks, and
ultimately a singular limit in which stress propagates along straight rays.
Recent experiments, as well as aspects of hyperbolic models, are discussed
in light of these results.
\end{abstract}

\begin{multicols}{2}

Descriptions of static granular matter as an isotropic elasto-plastic
continuum~\cite{reviews} do not incorporate effects of granularity essential
to reproduce the variety of stress profiles identified experimentally~\cite
{liu,vanel,silva,reydellet,serero,geng,geng2001,mueggenburg}. By contrast,
recent discrete models relying on (probabilistic) rules of force
transmission among individual grains \cite{liu,coppersmith,claudin1998}
require, for a continuum limit in terms of stresses, an additional
constraint relating the components of the stress tensor. This closure
relation may be interpreted~\cite{claudin1998} as a local Janssen
approximation~\cite{janssen,reviews} but is otherwise rather {\it ad hoc}.
Furthermore, it leads to hyperbolic equations for stress propagation \cite
{wittmer1996,wittmer1997,claudin1998}, as opposed to the usual elliptic
equations of elastic theories.

Reference~\cite{gay} offers a possible bridging approach in which a
continuum theory is derived systematically from grain-grain interactions,
taking into account the texture of the packing and the freedom of grains to
reorient with respect to their surrounding medium. In the present Letter, we
apply this general framework to calculate the propagation of the stress from
a point force applied to a two dimensional granular system. After recalling
the main elements of the theory and discussing its general properties in two
dimensions, we calculate stress profiles in the presence of a friction large
enough to prevent slip at the contacts. We obtain a variety of responses
that depend on the type of packing and provide us with a natural
interpretation of recent experiments. Finally, we discuss the case of easily
sheared contacts, which mimic slipping contacts and induce sharper stress
peaks. In the limiting situation, the hyperbolic closure relations used in
the literature~\cite{wittmer1996,wittmer1997,claudin1998} emerge from our
formulation and constrain the applied point force to propagate along
straight rays.

As shown in Ref.~\cite{gay}, a strain $\varepsilon $ about a preloaded
equilibrium state induces reorientation of the grains, expressed by the
linearized rotation matrix $\delta _{ij}+\Omega _{ij}$, and generates
incremental stresses 
\begin{equation}
\sigma _{ij}=(\varepsilon _{ik}-\Omega _{ik})\,Q_{kj}+P_{ijkl}\,\varepsilon
_{kl}.  \label{stress}
\end{equation}%
The fabric tensors%
\begin{equation}
Q_{kj}=\int k_{\bot }D_{k}D_{j}\mu \;d\alpha ,  \label{tensorQ}
\end{equation}%
\begin{equation}
P_{ijkl}=\int (k_{\Vert }-k_{\bot })\frac{D_{i}D_{j}D_{k}D_{l}}{D^{2}}\mu
\;d\alpha ,  \label{tensorP}
\end{equation}%
encode the texture of the granular packing {\it via} the average number of
inter-grain contacts $\mu (\alpha )d\alpha $ within an angle $d\alpha $
about $\alpha $, the normal and shear moduli $k_{\Vert }(\alpha )$ and $%
k_{\bot }(\alpha )$ of these contacts, and the average center-to-center
distances $D(\alpha )$ traversing them. In two dimensions $\Omega $ depends
on a single angle $\omega $, defined by
$\Omega =\pmatrix{0 & -\omega \cr \omega & 0}$ and set by the local
vanishing of torque to
\begin{equation}
\omega =\frac{(Q_{xx}-Q_{yy})\varepsilon_{xy}
+Q_{xy}(\varepsilon _{yy}-\varepsilon _{xx})}{Q_{xx}+Q_{yy}}
=\frac{[Q,\varepsilon]_{xy}}{{\rm Tr}(Q)}.  \label{angle}
\end{equation}
The stress then can be expressed in terms of the strain alone, for example
in the matrix form
\begin{equation}
\pmatrix{\sigma _{xx} \cr \sigma _{yy} \cr \sigma _{xy}}=M%
\pmatrix{\varepsilon _{xx} \cr \varepsilon _{yy} \cr 2\varepsilon _{xy}},
\label{elasticity}
\end{equation}%
where the (symmetric) compliance matrix $M$ summarizes the textured
stiffness of the medium through the first few Fourier moments of the
quantities $\mu _{Q}(\alpha )=\,k_{\bot }(\alpha )\,D^{2}(\alpha )\mu
(\alpha )$ and $\mu _{P}(\alpha )=\left[ k_{\Vert }(\alpha )-k_{\bot
}(\alpha )\right] \,D^{2}(\alpha )\mu (\alpha )$ associated with the two
fabric tensors. Defining the Fourier moments by
\begin{eqnarray}
2\pi \mu _{Q}(\alpha ) &=&2\kappa _{0}+4\kappa _{1}\cos (2\alpha )+4\kappa
_{2}\sin (2\alpha )+\ldots ,  \nonumber \\
2\pi \mu _{P}(\alpha ) &=&2\bar{\kappa}_{0}+4\bar{\kappa}_{1}\cos (2\alpha
)+4\bar{\kappa}_{2}\sin (2\alpha ) \\
&&+16\bar{\kappa}_{3}\cos (4\alpha )+16\bar{\kappa}_{4}\sin (4\alpha
)+\ldots ,  \nonumber  \label{fourier}
\end{eqnarray}
we write $M$ explicitely, as
\begin{equation}
M=\pmatrix{ \tilde{\kappa}_{0}+\tilde{\kappa}_{1}+\tilde{\kappa}_{3} &
-\tilde{\kappa}_{3} & \tilde{\kappa}_{2}+\tilde{\kappa}_{4} \cr
-\tilde{\kappa}_{3} &
\tilde{\kappa}_{0}-\tilde{\kappa}_{1}+\tilde{\kappa}_{3} &
\tilde{\kappa}_{2}-\tilde{\kappa}_{4} \cr
\tilde{\kappa}_{2}+\tilde{\kappa}_{4} &
\tilde{\kappa}_{2}-\tilde{\kappa}_{4} &
-\tilde{\kappa}_{3}+\tilde{\kappa}_{5}},  \label{compliance}
\end{equation}%
where $\tilde{\kappa}_{0}=\kappa _{0}+\bar{\kappa}_{0}$, $\tilde{\kappa}%
_{1}=\kappa _{1}+\bar{\kappa}_{1}$, $\tilde{\kappa}_{2}=(\kappa _{2}+\bar{
\kappa}_{2})/2$, $\tilde{\kappa}_{3}=\bar{\kappa}_{3}-\bar{\kappa}%
_{0}/4-\kappa _{2}^{2}/2\kappa _{0}$, $\tilde{\kappa}_{4}=\bar{\kappa}%
_{4}+\kappa _{1}\kappa _{2}/2\kappa _{0}$, and $\tilde{\kappa}_{5}=\kappa
_{0}/2-(\kappa _{1}^{2}+\kappa _{2}^{2})/2\kappa _{0}$. The fore-aft
symmetry ($\alpha \rightarrow \alpha +\pi $) of the contact distribution
forbids dipolar moments, and the anisotropic part of $M$ is composed of the
quadrupolar moments $\kappa _{1}$, $\bar{\kappa}_{1}$, $\kappa _{2}$, $\bar{
\kappa}_{2}$ and octupolar moments $\bar{\kappa}_{3}$, $\bar{\kappa}_{4}$.

Before discussing stress profiles generated by Eq.~(\ref{elasticity}), we
comment on the properties of the compliance. The non-negativity of $\mu _{Q}$
and $\mu _{P}$ restricts the allowed domain of the multipolar components; $
\bar{\kappa}_{0}$, $\bar{\kappa}_{1}$ and $\bar{\kappa}_{3}$, for example,
are bound by $2(\bar{\kappa}_{1}/\bar{\kappa}_{0})^{2}-1\leq 4\bar{\kappa}%
_{3}/\bar{\kappa}_{0}\leq 1$. An intuitive derivation of these
(Cauchy-Schwartz) inequalities uses a representation of $\mu _{Q}$ and $\mu
_{P}$ as linear combinations of delta functions with non-negative weights.
The multipolar components of $\mu _{Q}$ and $\mu _{P}$ then are themselves
linear combinations of the multipolar components of the delta functions with
the same non-negative weights, hence the former interpolate the latter. For
a (fore-aft symmetric) pair of delta functions $\mu _{\alpha _{0}}(\alpha
)=\delta (\alpha -\alpha _{0})+\delta (\alpha -\alpha _{0}-\pi )$, $\bar{
\kappa}_{0}(\mu _{\alpha _{0}})=1$, $\bar{\kappa}_{1}(\mu _{\alpha
_{0}})=\cos (2\alpha _{0})$, $\bar{\kappa}_{3}(\mu _{\alpha _{0}})=\cos
(4\alpha _{0})/4$, and the bound $2(\bar{\kappa}_{1}/\bar{\kappa}%
_{0})^{2}-1\leq 4\bar{\kappa}_{3}/\bar{\kappa}_{0}$ is achieved. The choice $
\mu (\alpha )=\mu _{\alpha _{0}}(\alpha )$ corresponds to the rather
pathological limit of a granular medium made up of an array of independent,
parallel columns at an angle $\alpha _{0}$, and lies on the boundary of the
allowed domain. This domain is included entirely in the domain of stability
of $M$, in which stress resists any deformation (with a positive energy
cost). As long as friction is high enough to prevent slip at the contacts,
the boundaries of the domain of stability and the allowed domain meet only
at the columnar systems just mentioned. Finally, the form of $M$ in Eq.~(\ref
{compliance}) and the independence of $\tilde{\kappa}_{0},\dots ,\tilde{
\kappa}_{5}$ ensure that, within the allowed domain, the granular material
can take on the most general form of anisotropic elasticity.

The equilibrium conditions $\partial _{j}\,\sigma _{ij}=f_{i}\,\delta
(x)\delta (y)$, where $f_{i}$ is a point force applied at the origin, are
supplemented by the closure equation $\partial _{xx}\varepsilon
_{yy}+\partial _{yy}\varepsilon _{xx}=2\partial _{xy}\varepsilon _{xy}$.
This relation specifies the constraint on a strain tensor that derives from
an underlying two-component displacement vector field, and is re-expressed as
\begin{equation}
(\partial _{yy},\partial _{xx},-\partial _{xy})\,M^{-1}\pmatrix{\sigma _{xx}
\cr \sigma _{yy} \cr \sigma _{xy}}=0  \label{closure}
\end{equation}%
to close the system of equations governing stresses. A classic and elegant
solution makes use of the Airy function $\phi $, defined through its
derivatives by $\delta _{ij}\partial _{\kappa \kappa }\phi -\partial
_{i}\partial _{j}\phi =\sigma _{ij}$ so as to satisfy the equilibrium
conditions automatically. Thus one needs to solve only Eq. (\ref{closure}) for
$\phi $, respecting the boundary conditions at the origin and at infinity.
Taking advantage of the properties of analytic functions on the plane, one
can translate this fourth order differential equation for $\phi $ into a
fourth order algebraic equation \cite{greenbook}. Its roots determine
completely the angular modulation that factors the usual $1/r$ dependence of
two dimensional elastic stresses. In an orthotropic material (with two
perpendicular axes of symmetry), the quartic polynomial in question reduces
to a bi-quadratic one \cite{greenpaper}, and henceforth we restrict
ourselves to this case for the sake of simplicity. Specifically, we choose
the $x$- and \thinspace $y$-axes as axes of symmetry, so that $\kappa _{2}=
\bar{\kappa}_{2}=\bar{\kappa}_{4}=0$.

Following the program just outlined, we calculate the radial stress $\sigma
_{rr}$ and the reorientation angle $\omega $ analytically and plot their
angular dependence on Fig. 1 \cite{expansion}. While observed modulated (
{\it e.g.}, double-peaked) responses motivate hyperbolic models \cite
{claudin1998,wittmer1996,wittmer1997}, it was emphasized recently \cite
{goldenberg} that they arise also in anisotropic elastic systems \cite
{greenpaper,greenbook,ottopreprint}. Accordingly, for a weakly disordered
square lattice oriented at $45^{\circ }$ from the \thinspace $x$- and $y$
-axes ($\bar{\kappa}_{3}<0$), we find that the point force propagates
through the material as a shallow stress double-peak (middle bold curve in
Fig. 1(a)). As the packing is further disordered, the two peaks merge into a
single peak which ultimately coincides with the isotropic response (thin
solid curve). Alternatively, if contacts close to the vertical are slightly
favored ($\kappa _{1},\bar{\kappa}_{1}<0$) {\it e.g.}, by gravity, the two
peaks amalgamate into a single wide peak (top bold curve). If, however, the
preparation of the system favors contacts close to the horizontal ($\kappa
_{1},\bar{\kappa}_{1}>0$), the two peaks become more pronounced (bottom bold
curve) and the stress propagates mostly along two directions. (We note that
there is no trivial relation between the latter and possible preferred
directions in the packing.)

\begin{figure}
\vspace*{.2truecm}
\epsfxsize=8truecm
\centerline{ \epsfbox{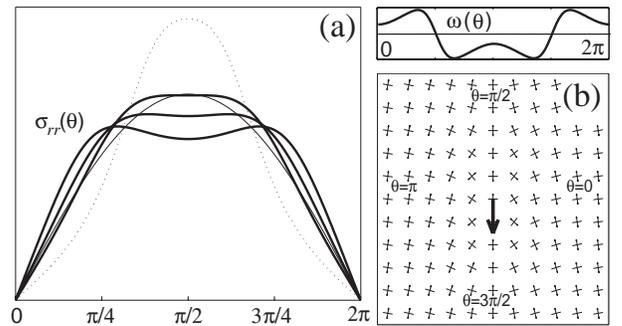} }

\caption{Radial stress and reorientation angle as a function of the polar
angle, for different packings of incompressible grains ($\protect\kappa _{1}=2\bar{
\protect\kappa}_{1}$ \protect\cite{johnson,gay}). (a) Bold curves: $
\bar{\protect\kappa}_{3}=-1/5$; $\bar{\protect\kappa}_{1}=-1/5$ (top), $\bar{
\protect\kappa}_{1}=0$ (middle), $\bar{\protect\kappa}_{1}=1/5$ (bottom).
Dotted curve: $\bar{\protect\kappa}_{1}=-7/10$, $\bar{\protect\kappa}_{3}=0$
. The isotropic result (thin solid line) is added for comparison. (b) $\bar{
\protect\kappa} _{1}=1/5$, $\bar{\protect\kappa}_{3}=-1/5$. The crosses
illustrate the angular dependence of the reorientation and its $1/r$ decay.}
\end{figure}

The plots are symmetric because they correspond to packings statistically
symmetric about the vertical (direction of the force). If
the material is rotated with respect to the force, one of the two peaks
dominates the other; more generally, if the force is not applied along an
axis of symmetry (as generically in a non-orthotropic medium), we expect a
biased response reminiscent of arching phenomena \cite{jenkin}. This is
certainly the case in the left (or right) half of a pile constructed with a
point source, where the packing is not symmetric about the vertical \cite
{geng2001,gay}. When $\kappa _{1}\neq 0$, the stress profile is due in part
to the reorientation of the grains. Figure 1(b) exhibits the angular
dependence of the reorientation angle $\omega $ corresponding to the
pronounced double-peak in Fig.~1(a); the crosses illustrate the form of the
reorientation field and its decay away from the origin. While the
reorientation appears to relieve the shear stress along the vertical as
expected intuitively, it shows a complicated structure and in particular
extrema along non-trivial directions.

In a remarkable measurement~\cite{geng} of the response of a vertical two
dimensional granular system (preloaded by its own weight) to a vertical
point force, the stress propagates along two peaks, in an ordered
(anisotropic) piling, which merge into a single peak as the piling is
disordered, in conformity with our picture. Furthermore, the width of the
single peak increases linearly with depth, in accordance with an elastic
theory. (The dependence of the widths on depth in the case of a double-peak
is not given.) Another experiment \cite{geng2001} correlates the macroscopic
response to the microscopic structure by recording both the stress profile
and the angular distribution of contacts $\mu (\alpha )$. As expected \cite
{gay}, $\mu (\alpha )$ shows a higher degree of anisotropy in a pile grown
with a point source than in a pile grown with an extended source, and
different stress profiles are measured: a double-peak for a point source, a
single peak for an extended source. In similar experiments on three
dimensional piles \cite{vanel,reydellet} the observed stress peak, while
scaling linearly in depth, is narrower than expected from isotropic
elasticity in an infinite half space \cite{reydellet}. The authors interpret
this accentuated response as possibly arising from the experimental boundary
conditions \cite{reydellet}---a conjecture left unconfirmed by a more
detailed study \cite{serero}. Our approach affords a complementary
explanation: no matter how one disorders the pile during or after its
growth, gravity sets a preferred direction; one then expects to have more
near-vertical contacts than near-horizontal ones, and a resulting narrower
response (dotted curve in Fig. 1(a)). This interpretation is supported by
measurements \cite{serero} showing that a dense packing yields a wider peak
than a loose packing. To go from dense to loose, one needs to
\textquotedblleft open holes\textquotedblright\ while maintaining stability;
thus these holes may decrease the number of near-horizontal contacts
relative to the number of near-vertical ones, but not the other way around.

To complete our picture of stress propagation, we discuss the degenerate
solutions that arise when slip occurs. As long as a large friction prevents
slip, the ratio of shear to normal moduli of the contacts is fixed~(for
purely compressively preloaded contacts, $k_{\bot }/k_{\Vert }=\left( 2-2\nu
\right) /\left( 2-\nu \right) $, where $\nu $ is the Poisson ratio of the
constitutive material of the grains~\cite{johnson}). At smaller frictions,
grains may slip along their contacts, effectively lowering the value of $
k_{\bot }$ \cite{makse}. This renormalization of the shear modulus yields
narrower stress peaks, and in the limit $k_{\bot }/k_{\Vert }\rightarrow 0$
singular responses arise in ordered packings. For a simple two dimensional
crystal (Bravais lattice) $\mu _{Q}(\alpha ),\mu _{P}(\alpha )\propto
\sum_{i=1}^{n}[\delta (\alpha -\alpha _{i})+\delta (\alpha -\alpha _{i}-\pi
)]$, where steric constraints impose $n\leq 3$~\cite{torquato}. The case $n=1
$ corresponds to the pathological columnar systems mentioned above, while $
n=3$ yields an isotropic elastic theory since the contact angles differ by
multiples of $\pi /3$. In the case $n=2$, the stress peaks become infinitely
narrow and coincide with the lattice directions $\alpha _{1}$ and $\alpha
_{2}$. Similar ray-like responses have been observed recently in three
dimensional crystalline packings \cite{mueggenburg}. In two dimensions, the
situation is easiest to visualize in a square lattice at $45^{\circ }$
(Fig.~2); a point force applied vertically is transmitted along two rows of
grains without disturbing any other grains.
We note that, in general, in the
limit $k_{\bot }/k_{\Vert }\rightarrow 0$ only the fabric tensor $P$
survives in Eq.~(\ref{stress}) and the system becomes equivalent to a
collection of fibers highly resistant to
compression and extension but easy to shear, as in much studied
fiber-reinforced materials \cite{pipkin}.
Ordering corresponds to aligning the fibers
along a set of given directions;
in the case of only two directions
there is at least one soft mode,
associated with a singular response. In particular, the conformation
of Fig.~2 is equivalent to two sets of fibers at right angle,
equivalent in turn to an incompressible elastic medium with fibers
along a single direction.

\begin{figure}
\vspace*{-.1truecm}
\epsfxsize=8truecm
\centerline{ \epsfbox{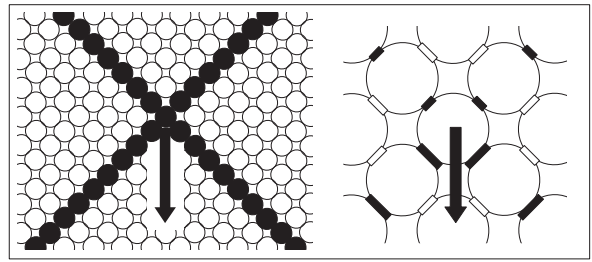} }

\caption{Illustration of a singular response in an ordered packing. Left:
the force is transmitted through the two rows of black grains. Right:
detail; over- and under-compressed contacts are in black, unperturbed ones
in white.}
\vspace*{-.1truecm}
\end{figure}

Singular solutions of a hyperbolic character occur only at the boundary of
the domain of stability of the elastic theory. In the limit $k_{\bot
}/k_{\Vert }\rightarrow 0$, this boundary and that of the allowed domain
meet not only at the $n=1$ columnar systems but also at the $n=2$
crystalline packings. For a packing symmetric with respect to the $x$
- and $y$-axes ($\alpha _{1}=-\alpha _{2}\equiv \alpha _{0}$), the closure
Eq.~(\ref{closure}) reduces to%
\begin{equation}
(\partial _{xx}-c_{0}^{-2}\partial _{yy})(\sigma _{xx}-c_{0}^{2}\,\sigma
_{yy})=0,  \label{hyperbolicclosure}
\end{equation}%
where $c_{0}^{-2}=(\bar{\kappa}_{0}-\bar{\kappa}_{1})/(\bar{\kappa}_{0}/4-
\bar{\kappa}_{3})-1=\tan ^{2}\alpha _{0}\geq 0$. If the system is rotated by
an angle $\arctan \left( t\right) $, the closure equation involves the more
general linear combination $\left( 1-c_{0}^{2}t^{2}\right) \sigma
_{xx}-\left( c_{0}^{2}-t^{2}\right) \sigma _{yy}-2(1+c_{0}^{2})\,t\sigma
_{xy}$.
Since the limit $k_{\bot }/k_{\Vert }\rightarrow 0$ may be viewed also as
that of infinitely hard grains (with a thin incompressible elastic coating),
it is no surprise that we retrieve the form of the closure equations derived 
\cite{edwards1999,ball} for the marginal (isostatic) state of a granular
packing in which the response may be calculated without knowledge of
deformations. In integral form, our closure equation with $c_{0}^{2}=1$ is
identical to the one given in Ref. \cite{edwards1999}, while for general $
c_{0}$ and $t$ it coincides with that of Ref. \cite{ball} and the
phenomenological \textquotedblleft oriented stress linearity
model\textquotedblright\ of Ref. \cite{wittmer1997}. In Ref. \cite
{claudin1998}, an integral form of Eq. (\ref{hyperbolicclosure}) is
suggested by a discrete probabilistic model and the packing disorder is
emulated by adding to $c_{0}^{2}$ a small increment that fluctuates randomly
in space. This uncertainty in the local propagation directions results in a
diffusive broadening of the stress rays, proportional to the square root of
the depth. Within our approach, by contrast, as soon as disorder in the
contact orientation is introduced, the response of a homogeneous system is
governed by elastic (elliptic) equations on large scales, yielding a linear
broadening of the stress peaks. These remarks help conciliate the
square-root broadening observed \cite{silva} in a pile less than ten grains
in height with the linear broadening \cite{reydellet,serero,geng} on larger
scales. The situation might be different, however, for heterogeneous media;
in a polycrystalline packing, for example, one might expect a diffusive
behavior even on macroscopic scales.

It is a pleasure to thank D.S. Fisher and J.R. Rice, as well as J.-P.
Bouchaud, P.-G. de Gennes, B.I. Halperin, and T.A. Witten, for inspiring
discussions. RdS gratefully acknowledges the support of the Harvard Society
of Fellows and the Milton Fund.

\end{multicols}


\begin{thebibliography}{99}
\bibitem{expansion} The results shown in Fig. 1 are derived at strong
anisotropy, for which the approach of Ref. \cite{gay} is best suited. At
weak anisotropy, $\bar{\kappa}_{1}$ and $\bar{\kappa}_{3}$ correct the
isotropic response according to
\begin{equation}
\begin{array}{lll}
\frac{2\pi r}{17/11}\sigma _{rr} & = & f_{x}\cos \theta +f_{y}\sin \theta 
\\ 
&  & \frac{48}{187}\left\{ 3\bar{\kappa}_{1}\left[ f_{x}\cos \theta \left( -
\frac{7}{24}+\cos 2\theta \right) \right. \right.  \\ 
&  & +\left. f_{y}\sin \theta \left( \frac{7}{24}+\cos 2\theta \right)
\right]  \\ 
&  & +\bar{\kappa}_{3}\left[ f_{x}\cos \theta \left( -\frac{7}{15}+\cos
2\theta +\frac{3}{5}\cos 4\theta \right) \right.  \\ 
&  & \left. \left. f_{y}\sin \theta \left( -\frac{7}{15}-\cos 2\theta +\frac{
3}{5}\cos 4\theta \right) \right] \right\} ,
\end{array}
\end{equation}
where we neglected higher order terms and assumed incompressible grains 
($k_{\Vert}/k_{\bot }=2/3$, $\kappa _{1}=2\bar{\kappa}_{1}$ \cite{johnson,gay}).
\end{thebibliography}
\end{document}